\pdfoutput=1

\documentclass[10pt,journal]{IEEEtran}
\ifCLASSOPTIONcompsoc
  \usepackage[nocompress]{cite}
\else
  \usepackage{cite}
\fi

\newtheorem{theorem}{Theorem}

\newtheorem{assumptions}{Assumptions}
\newtheorem{notation}{Notation}

\usepackage{graphicx}
\usepackage{subfig}
\usepackage{enumitem}
\usepackage{amsmath}
\usepackage{adjustbox}
\usepackage{amsfonts}
\usepackage{amssymb}

\begin{document}

\title{Towards the Mathematical Foundation of the Minimum Enclosing Ball and Related Problems}

\author{Michael~N.~Vrahatis
\IEEEcompsocitemizethanks{\IEEEcompsocthanksitem M.~N.~Vrahatis is with the Department
of Mathematics, University of Patras, GR-26110 Patras, Greece.\protect\\
E-mail: vrahatis@math.upatras.gr\,\, \&\,
vrahatis@upatras.gr}
}
\IEEEtitleabstractindextext{
\begin{abstract}
Theoretical background is provided towards the mathematical foundation of the minimum enclosing ball problem.
This problem concerns the determination of the
unique spherical surface of smallest radius enclosing a given bounded set
in the \emph{d}-dimensional Euclidean space.
The study of several problems that are similar or related to the minimum enclosing ball problem
has received a considerable impetus from the large amount of applications of these problems in various fields of science and technology.
The proposed theoretical framework is based on several enclosing (covering) and partitioning (clustering) theorems and provides among others bounds and
relations between the circumradius, inradius, diameter and width of a set.
These enclosing and partitioning theorems
are considered as
cornerstones in the field that strongly influencing
developments and generalizations to other spaces and non-Euclidean geometries.
\end{abstract}

\thispagestyle{empty}

}
\maketitle

\IEEEdisplaynontitleabstractindextext

\IEEEpeerreviewmaketitle

\ifCLASSOPTIONcompsoc
\IEEEraisesectionheading{\section{Introduction}\label{sec:introduction}}
\else
\section{Introduction}\label{sec:introduction}
\fi

\IEEEPARstart{T}{heoretical} background is provided that is
related to the \emph{Minimum Enclosing Ball (MEB)}\/ problem,
which refers to the determination of
the unique spherical surface of
smallest radius enclosing a given bounded subset of
the $d$-dimensional Euclidean space $\mathbb{R}^d$ (also denoted as~$\mathbb{E}^d$).
This problem is also referred in the literature using, among others, the following nomenclatures
(appearing in alphabetical order):
(1)~\emph{Chebyshev radius and Chebyshev center},
(2)~\emph{$d$-outer radius},
(3)~\emph{Euclidean 1-center},
(4)~\emph{minimax problem in facility locations},
(5)~\emph{minimum bounding sphere},
(6)~\emph{minimum covering sphere},
(7)~\emph{minimum enclosing sphere},
(8)~\emph{minimum spanning ball},
(9)~\emph{smallest bounding ball},
(10)~\emph{smallest circle problem},
(11)~\emph{smallest covering sphere},
(12)~\emph{smallest enclosing ball}, and
(13)~\emph{smallest enclosing sphere}.
It is worth noting that,
these nomenclatures, sometimes, appear slightly different in several fields of science and technology.
For examble, they appear as:
(a)~\emph{1-center problem}\/ in computational geometry,
(b)~\emph{1-point estimator}\/ in computational statistics,
(c)~\emph{1-class classification}\/ in machine learning, and
(d)~\emph{minimax optimization problem}\/ in operations research and optimization.
We note in passing that, according to our experience, the nomenclature used most frequently worldwide
is the \emph{{minimum enclosing ball problem}}, MEB problem in short.

It is worth mentioning here that,
for studying and analysing MEB and related to it problems four main
characteristics of a set $P$ in $\mathbb{R}^d$ can be used, among others.
Specifically:
(a)
The \emph{circumradius} of~$P$, \emph{i.e}.\ the radius of the smallest spherical surface enclosing $P$,
(b)
The \emph{inradius} of~$P$, \emph{i.e}.\ the radius of the greatest spherical surface which is contained in $P$,
(c)
The \emph{diameter} of~$P$, \emph{i.e}.\ the maximal distance of any two points of $P$,  and
(d)
The \emph{width} of~$P$, \emph{i.e}.\
the minimal distance between a pair of parallel supporting hyperplanes (tac-hyperplanes) of~$P$.

It is believed that the earliest known statement of the MEB problem was first posed in \emph{circa}\/ 300 B.C.
by the considered ``father of geometry''
Euclid\footnote{Euclid (\emph{c.}~323 -- \emph{c.} 285 B.C.), Greek mathematician.}
in his seminal work \emph{``Elements''} and specifically in Book IV, Proposition 5
which refers to circumscribe a circle about a given triangle~\cite{Euclid1482}.
The general case of finding the smallest circle enclosing a given finite set of $n$ points in $\mathbb{R}^2$
was first appeared in 1857 by
Sylvester\footnote{James Joseph Sylvester (1814 -- 1897), English mathematician.}
\cite{Sylvester1857}.
In addition, the earliest known method for tackling the above case was also proposed in 1860 by Sylvester~\cite{Sylvester1860}.
The same method was independently proposed in 1885 by
Chrystal\footnote{George Chrystal (1851 -- 1911), Scottish mathematician.}
\cite{Chrystal1885}.
Thereafter, various methods for determining the MEB of a set of~$n$ points in the plane
have been proposed.

In 1901
Jung\footnote{Heinrich Wilhelm Ewald Jung (1876 -- 1953), German mathematician.}
in his dissertation established seminal results
regarding the MEB problem~\cite{Jung1901}.
Thus, Jung was first answered to the question of best possible estimate of the radius of a sphere
in $\mathbb{R}^d$ enclosing a bounded subset of $\mathbb{R}^d$
of a given diameter
(\emph{cf}. {Jung's Theorem} \ref{JungTh}).
Thereafter, generalizations of MEB problem to $d$-dimensions
have been of interest for many years and this problem remains an important issue of study and further research in the field.
In general, \emph{Jung's enclosing theorem}\/ provides an upper bound of the circumradius of a set in $\mathbb{R}^d$ in terms of its diameter.
A similar important theorem that it has been proposed by
Steinhagen\footnote{Paul Steinhagen, German mathematician, Dr. phil. Univ. Hamburg 1920.}
in 1921~\cite{Steinhagen1921}
provides lower bounds (for $d$ odd and $d$ even) of the inradius of a set in terms of its width
(\emph{cf}.\ Steinhagen's Theorem~\ref{SteinhagenTh}).

The MEB problem can be tackled by applying \emph{enclosing (covering) theorems}\/
and it is, among others,  one of the most fundamental problems in {clustering}.
Specifically, \emph{clustering}\/
with respect to the \emph{diameter}\/ and the \emph{radius} costs, is the task of
\emph{partitioning}\/ a set of points in $\mathbb{R}^d$
to subsets,
where items in the same subset, named \emph{cluster},
are similar to each other, compared to items in other clusters~\cite{AlonDPR2004}.
Usually, clustering problems arise in the analysis of large data sets.
Thus, the \emph{approximate clustering via core-sets} is used
for clustering of a set of points in $\mathbb{R}^d$ (for large $d$\,) by
extracting properly a small set of points named \emph{core-set}\/
that ``represents'' the given set of points~\cite{BadoiuHI2002}.
These issues lie within the domain of {property testing}.
The \emph{property testing}\/
is motivated for designing and development of ``super-fast'' algorithms for analysing the
global structure of datasets that are too large to read in their entirety in a reasonable
time.
These algorithms have direct access to items of a
``huge'' input data set and determine whether this data set satisfies
a desired (predetermined global) property, or is ``far'' from satisfying it.
They inspect relatively small portions of the data set and
their complexity is measured in terms of the number of accesses to the input.
The algorithms that are used are necessarily \emph{randomized}
(otherwise they may be drawing a conclusion from an atypical portion of the
input) and \emph{approximate} (since it is not expected the algorithm to produce an exact
answer having examined only a portion of the input)
(\emph{cf}.\ \cite{Goldreich2017}, \cite{ChakrabortyPRS2018}).
In these directions and trends a significant role is played by
\emph{partitioning theorems}, \emph{i.e}.\ Helly-type theorems~\cite{BaranyK2022},
that constitute fundamental results describing
the ways in which convex sets intersect with each other.
For an illustrative interesting application of partitioning theorems
in property testing for solving the clustering problem we refer the interested reader to~\cite{ChakrabortyPRS2018}.

The paper at hand aims to contribute towards the mathematical foundation of a
framework to analyze and study rigorously the MEB
problem and the similar and related to it issues and aspects.
Therefore, theoretical results are summarized in a number of significant
enclosing (covering) and partitioning (clustering) theorems that can be utilized for tackling
the MEB and the related to it problems.
To this end, the rest of the paper,
after Section~\ref{sec:Prelim} where a preliminary background material is provided,
focuses on the following issues and aspects.
In Section~\ref{sec:RelAppl}
several problems that are related to the MEB  problem and various of their applications are shorty presented.
Then, the following issues are given that forms the proposed framework.
Specifically, in Section~\ref{sec:Relat} all possible inequalities between any two of
circumradius, inradius, diameter and width of a set in $\mathbb{R}^d$ are given.
Also, various interesting  properties of the regular simplex that are useful for the MEB problem are highlighted.
In addition, bounds for the quotient of inner and outer radii of a compact (\emph{i.e}.\ closed and bounded) convex body
are presented and the determination of the inner and outer radii of three
regular polytopes namely, the
regular simplex, hypercube, and regular cross-polytope are provided.
In Section~\ref{sec:EnclTh} several enclosing and partitioning theorems are provided
including, among others, the theorems of Jung, Steinhagen, Perel${}^\prime\!$man,
Carath\'{e}odory, Helly and Tverberg.
Also, various generalizations to other spaces
and non-Euclidean geometries, mainly of Jung's
theorem are highlighted.
The paper ends in Section~\ref{sec:Synop} with concluding remarks.

\section{Preliminary Background Material}\label{sec:Prelim}

The following notions are basic to the analysis and study
of the MEB problem and the similar and related to it problems.

The straightforward generalization of the concept of a triangle to the $d$\!\! -\!\! dimensional Euclidean space $\mathbb{R}^d$ is the
\emph{$d$-simplex} (see, \emph{e.g.}\ \cite{Greenberg1967}).
In general, the $d$-simplex is considered to be as the smallest possible $d$-dimensional polytope,
\emph{i.e.}\ the convex hull of a finite number of points in $\mathbb{R}^d$~(\emph{cf}.\ \cite[p.\ 8]{Guggenheimer1977}).

Formally, for a set of points $\{ {\upsilon}^k\}_{k=0}^d$ in $\mathbb{R}^d$
that are affinely independent (\emph{i.e}.\ the vectors
$\{ {\upsilon}^k - {\upsilon}^0\}_{k=1}^d$ are linearly independent)
the convex hull
${\rm conv}\{{\upsilon}^0,$ $ {\upsilon}^1, \ldots,$ ${\upsilon}^d \} \equiv
[ {\upsilon}^0,$ ${\upsilon}^1, \ldots,$ ${\upsilon}^d ]$
is called the {\em $d$-simplex with vertices}
${\upsilon}^0,$ ${\upsilon}^1, \ldots, {\upsilon}^d$
and is denoted by
$\sigma^d = [{\upsilon}^0, {\upsilon}^1, \ldots,{\upsilon}^d]$
while its boundary is denoted by ${\vartheta} {\sigma}^d$.
The order of the points $\{{\upsilon}^k\}_{k=0}^d$ that are written in a list determines an orientation.
On the other hand, since the convex hull is independent of this order the vertices can be permuted in a list.

A $d$-simplex is called \emph{oriented}\/ if an order has been assigned to its vertices.
If $\langle {\upsilon }^0, {\upsilon }^1, \ldots ,{\upsilon }^d \rangle $ is an orientation of
$\{ {\upsilon }^0, {\upsilon }^1, \ldots ,{\upsilon }^d \} $
this is considered as being the same as any orientation that can be attained from it by an even
permutation of the vertices and as the opposite of any orientation
attained by an odd permutation of the vertices.
The oriented $d$-simplex is denoted by
${\sigma }^d =
\langle {\upsilon}^0, {\upsilon }^1, \ldots ,{\upsilon }^d \rangle, $
and it can be denoted, for example,
$\langle {\upsilon }^0, {\upsilon }^1, {\upsilon }^2,
 \ldots ,{\upsilon }^d \rangle =
- \langle {\upsilon }^1, {\upsilon }^0, {\upsilon }^2,
\ldots ,{\upsilon }^d \rangle =
\langle {\upsilon }^2, {\upsilon }^0, {\upsilon }^1,
 \ldots ,{\upsilon }^d \rangle .$
The \emph{boundary}\/ ${\vartheta} {\sigma}^d$ of an oriented
 $d$-simplex ${\sigma}^d =
\langle {\upsilon }^0, {\upsilon }^1, \ldots,$ ${\upsilon }^d \rangle $
 is given by
 $
 {\vartheta} {\sigma}^d =
   \sum_{i=0}^{d} { (-1)^i
\langle {\upsilon }^0, {\upsilon }^1, \ldots ,
{\upsilon }^{i-1}, {\upsilon }^{i+1}, \ldots ,
{\upsilon }^d \rangle }.
$
A $d$-dimensional \emph{polyhedron} ${\mathit\Pi}^d$ is a union of a
finite number $k$\/ of oriented $d$-simplices ${\sigma}^d_{(t)}$,
$t = 1,2,\ldots,k$ such that the ${\sigma}^d_{(t)}$ have
pairwise-disjoint interiors. Thus, it can be written
${\mathit\Pi}^d = \sum_{t=1}^{k} {\sigma}^d_{(t)}$
and
${\vartheta}{\mathit\Pi}^d = \sum_{i=1}^{k} {\vartheta}{\sigma}^d_{(t)}.$

The convex hull of every nonempty subset of the $d + 1$ points that determine
a $d$-simplex is called a \emph{face}\/ of the simplex.
The faces are simplices themselves and they are also called {\em facets}.
Specifically, for each subset of $m+1$ elements
$\{ u^0, u^1, \ldots , u^m \} \subset
\{ {\upsilon}^0, {\upsilon}^1, \ldots ,{\upsilon}^d \}, $
the $m$-simplex
$[ u^0, u^1, \ldots , u^m ]$
is called an {\em $m$-face} of
$[ {\upsilon}^0,$ $ {\upsilon}^1, \ldots,{\upsilon}^d ].$
In particular, 0-faces are  vertices and 1-faces are edges.
If all the edges have the same length, the simplex is called \emph{regular}\/
or \emph{equilateral}\/ (\emph{cf.}\ \cite{BlumenthalW1941}).
The $(m-1)$-simplex that defines
the $i$-th $(m-1)$-face opposite to the vertex ${\upsilon}^i$ of ${\sigma}^m$
is denoted by
$\sigma^{m}_{\neg i} = [{\upsilon}^0,$ $
{\upsilon}^1, \ldots,$ ${\upsilon}^{i-1}, {\upsilon}^{i+1}, \ldots,
{\upsilon}^m]$.
Similarly,
the $(m-2)$-simplex that determines the $j$-th $(m-2)$-face opposite to
the vertex ${\upsilon}^j$ of $\sigma^{m}_{\neg i}$,
for $j \neq i$, is denoted by
$\sigma^{m}_{\neg ij} = [{\upsilon}^0,$ $
{\upsilon}^1, \ldots,$ ${\upsilon}^{i-1}, {\upsilon}^{i+1}, \ldots,$
${\upsilon}^{j-1}, {\upsilon}^{j+1}, \ldots,
{\upsilon}^m]$.
The boundary of $\sigma^m$ in terms of its $(m-1)$-faces is given by
${\vartheta} {\sigma}^m = \bigcup _{i=0}^{m} \sigma^{m}_{\neg i}$.
The number of the $k$-faces of $\sigma^m$
is equal to the binomial coefficient $\binom{m+1}{k+1}$ (\emph{cf.} \cite[p.\ 120]{Coxeter1948}).
For instance, for $k=0$ the number of 0-faces (vertices)
is $\binom{m+1}{1}= m+1$, in addition for $k=1$ the number of 1-faces (edges)
is $\binom{m+1}{2}= m(m+1)/2$, while for $k=m-1$
the number of $(m-1)$-faces is $\binom{m+1}{m}= m+1$.

The \emph{diameter}\/ of
an $m$-simplex $\sigma^m$ in $\mathbb{R}^d$ ($d \geqslant m$) denoted by ${\rm diam}(\sigma^m)$,
is the length of the longest edge (1-face) of $\sigma^m$
(\emph{cf.}\ \cite[p.\ 812]{Whitehead1940}, \cite[p.\ 607]{AlexandroffH1974}),
where the Euclidean norm, $\|\cdot\|$, is used here to measure distances.
The length of the shortest edge of $\sigma^m$ will be denoted by ${\rm shor}(\sigma^m)$.
The \emph{width}\/ or \emph{breadth} of $\sigma^m$ denoted by ${\rm wid}(\sigma^m)$, is
the minimum distance between a pair of
parallel supporting hyperplanes (tac-hyperplanes) of~$\sigma^m$.

The \emph{barycenter}\/ or \emph{centroid}\/ of
an $m$-simplex in $\mathbb{R}^d$ ($d \geqslant m$)
$\sigma^m = [{\upsilon}^0, {\upsilon}^1, \ldots ,{\upsilon}^m ]$
and the \emph{barycenter}\/ or \emph{centroid}\/ of the
$i$-th $(m-1)$-face $\sigma^{m}_{\neg i}$ for $i=0, 1, \ldots, m$ of $\sigma^m$
are respectively denoted by $\kappa^m$ and $\kappa_i^m$ and are given by
$\kappa^m = (m+1)^{-1} \sum_{j=0}^{m} {\upsilon}^j$\,
and\,
$\kappa_i^m  = m^{-1} \sum_{j=0,\,j\ne i}^m {\upsilon}^j.$
By convexity the barycenter of $\sigma^m$
is a point in the relative interior of $\sigma ^m$.
The 1-simplex ${\mu}_i^m =\bigl[{\upsilon}^i,\, \kappa_i^m\bigr]$ for $i=0, 1, \ldots, m$ is called the \emph{$i$-th median}\/ of~$\sigma^m$
that corresponds to the vertex ${\upsilon}^i$.
The $m+1$ medians of
$\sigma^m$ concur in the barycenter $\kappa^m$ of $\sigma^m$
that divides each of them in the ratio $1:m$, where the longer segment being on
the side of the vertex of $\sigma^m$ (\emph{cf}.\ Theorem~\ref{SimCommandinoTh}).
The $\sigma^m$ is called \emph{orthocentric}\/ if its $m+1$ altitudes intersect in a common point $o^m$, called its \emph{orthocenter}.
It should be noted that
the $m+1$ altitudes of an $m$-simplex are not necessarily concurrent
if $m \geqslant 3$.

The $(d-1)$-dimensional spherical surface is denoted by ${S}^{d-1}$ for which ${S}^{d-1} = \vartheta {B}^d_{r,c}$
where ${B}^d_{r,c} \subset \mathbb{R}^d$ denotes the standard closed Euclidean ball of radius $r > 0$ centered at a point $c$ in $\mathbb{R}^d$,
\emph{i.e.}\ ${B}^d_{r,c} = \{ x \in \mathbb{R}^d : \| x -c \| \leqslant r \}$.
The smallest radius of the spherical surfaces ${S}^{d-1}$
that enclose an $m$-simplex $\sigma^m$ in\/ ${\mathbb{R}}^d$ $(d \geqslant m)$  is called \emph{circumradius}\/
and it is denoted by $\rho_{\rm cir}^m (\sigma^m)$ and
the corresponding center denoted by $c_{\rm cir}^m$ is called \emph{circumcenter}.
If $c_{\rm cir}^m = \kappa^m$, where $\kappa^m$ is the barycenter of $\sigma^m$ then the corresponding circumradius
is called \emph{barycentric circumradius}\/ and is denoted by $\beta_{\rm cir}^m (\sigma^m)$.
The largest radius of the spherical surfaces ${S}^{d-1}$
that are enclosed within $\sigma^m$ is called \emph{inradius}\/ of $\sigma^m$, it is denoted by $\rho_{\rm inr}^m (\sigma^m)$ and
the corresponding center denoted by $c_{\rm inr}^m$ is called \emph{incenter}.
If $c_{\rm inr}^m = \kappa^m$ then the corresponding inradius denoted by $\beta_{\rm inr}^m (\sigma^m)$ is called \emph{barycentric inradius} and it
is given~by
$\beta_{\rm inr}^m (\sigma^m) = \min_{x \in {\vartheta}{\sigma}^m} \|\kappa^{m} - x\|$.
In addition, the quantity
${\theta}({\sigma}^m) = \beta_{\rm inr}^m (\sigma^m) / {\rm diam}(\sigma^m),$
is called \emph{thickness}\/ of $\sigma^m$
that provides a measure for the quality or how well shaped a simplex is (\emph{cf.}\ \cite{Whitehead1940},\cite{Vrahatis2023}).

\section{Uniqueness, Related Problems, Theory and Applications of the MEB Problem}\label{sec:RelAppl}

The solution of the MEB problem has been proved that is \emph{unique}\/ in the following cases:
\begin{itemize}
\leftskip0.05cm
\item[(1)]
The \emph{Euclidean geometry} (see, \emph{e.g.} Welzl 1991 \cite{Welzl1991}).
\item[(2)] The \emph{hyperbolic geometry} (see, \emph{e.g.} Nielsen and Hadjeres 2015 \cite{NielsenH2015}).
\item[(3)] The \emph{Riemannian positive-\! definite matrix manifold} (see, \emph{e.g.} Lang 1999 \cite{Lang1999}, Nielsen and Bhatia 2013 \cite{NielsenB2013}).
\item[(4)] In any \emph{Cartan-Hadamard manifold} (see, \emph{e.g.} Arnaudon and Nielsen 2013 \cite{ArnaudonN2013}),
\emph{i.e.} Riemannian manifold that is complete and simply connected with non-\!\! positive sectional curvatures (see, \emph{e.g.} Nielsen 2020 \cite{Nielsen2020}).
\item[(5)] In any
\emph{Bruhat-Tits space} (see, \emph{e.g.} Lang 1999 \cite{Lang1999}),
that is complete metric space with a semi-parallelogram property,
which includes the Riemannian manifold of symmetric positive definite matrices (see, \emph{e.g.} Nielsen 2020~\cite{Nielsen2020}).
\end{itemize}
On the other hand, the solution of the MEB problem may not be unique in a metric space,
as for instance, in the case of \emph{discrete Hamming metric space} (see, \emph{e.g.} Mazumdar \emph{et al.} 2013 \cite{MazumdarPS2013}) which,
in this case, results to NP-hard computation (see, \emph{e.g.} Nielsen 2020 \cite{Nielsen2020}).

Several problems are related or similar to the MEB problem including, among others, the following ones:
\begin{itemize}
\leftskip0.05cm
\item[(1)] The computation of the \emph{minimum number of unit disks on a plane}\/  required to cover the $n$ points of a set in $\mathbb{R}^2$, along with a
placement of the disks, (see, \emph{e.g.} Friederich \emph{et al.} 2023 \cite{FriederichGGHS2023}).
\item[(2)] The determination of the \emph{smallest enclosing sphere of a finite number of nonempty closed subsets of $\mathbb{R}^d$},
as well as the closely related problem of determining a sphere with the smallest radius that intersects all of the sets
(see, \emph{e.g.} Mordukhovich \emph{et al.} 2013 \cite{MordukhovichNV2013}).
\item[(3)] The determination of the \emph{smallest sphere that covers at least $(1 - \gamma ) n$ points of a set}\/ with $n$ points in $\mathbb{R}^d$,
where $\gamma \in (0, 1)$ is a parameter that it could be named \emph{outliers exclusion parameter}\/
(see, \emph{e.g.} Ding 2021 \cite{Ding2021}).
\item[(4)]
The determination of the \emph{minimum $k$-enclosing ball problem}\/ that consists of finding
the ball with smallest radius that contains at least $k$ of $n$ given points of a set in $\mathbb{R}^d$
(this case is similar to the previous one)
(see, \emph{e.g.} Cavaleiro and Alizadeh 2022 \cite{CavaleiroA2022}).
\item[(5)]
The determination of the \emph{complexity of the minimum $k$-enclosing ball problem}\/
(see, \emph{e.g.} Shenmaier 2013 \cite{Shenmaier2013}).
\item[(6)] The identification and elimination of the \emph{interior points}\/ of the minimum radius ball enclosing a set in $\mathbb{R}^d$
(see, \emph{e.g.} Ahipa\c{s}ao\u{g}lu and Yıldırım 2008 \cite{AhipasaogluY2008}).
\item[(7)] The determination of the \emph{smallest enclosing ball of a given set of $n$ balls}\/ in $\mathbb{R}^d$
(see, \emph{e.g.} Fischer and G\"{a}rtner 2004 \cite{FischerG2004}).
\item[(8)]
The treatment of the \emph{Euclidean $k$-median}\/ problem that aims at
the localization of $k$ medians (facilities) among the given $n$ points of a set in $\mathbb{R}^d$
such that the sum of the distances from each point of the set to its closest median to be minimized
(see, \emph{e.g.} Kolliopoulos and Rao 2007 \cite{KolliopoulosR2007},
Bhattacharya \emph{et al.} 2021 \cite{BhattacharyaGJ2021}).
\item[(9)] The treatment of
the \emph{uncapacitated facility location}\/ problem that requires locating an undetermined
number of facilities to minimize the sum of the
fixed setup costs and the variable costs of serving the market demand from these facilities.
In several cases of real life applications capacity limitations are incorporated on the facilities
to be established. This problem is known as \emph{capacitated facility location}\/ problem
(see, \emph{e.g.} Verter 2011 \cite{Verter2011}, Kolliopoulos and Moysoglou 2016 \cite{KolliopoulosM2016}).
\item[(10)]
The determination of a given point in $\mathbb{R}^d$ its \emph{$k$-nearest neighbor points of a set}\/ in $\mathbb{R}^d$
(see, \emph{e.g.} Zhang, 2022~\cite{Zhang2022}, Syriopoulos \emph{et al.} 2023 \cite{SyriopoulosKKV2023}).
\item[(11)]
The treatment of the
\emph{nearest-neighbor-preserving embeddings}\/ that are randomized embeddings between two metric spaces which preserve
the approximate nearest-neighbors
(see, \emph{e.g.} Indyk and Naor 2007~\cite{IndykN2007}, Emiris \emph{et al.} 2023 \cite{EmirisMP2023}).
\item[(12)]
The treatment of the \emph{$k$-center problem}\/ that aims at
the determination of $k$ balls with the smallest radius to cover a finite number of given points in $\mathbb{R}^d$.
A \emph{generalized version of the $k$-center problem}\/ that aims at finding $k$ balls with the smallest radius
such that their union intersects all the elements of a given finite collection of nonempty closed
convex sets in $\mathbb{R}^d$
(see, \emph{e.g.} An \emph{et al.} 2020 \cite{AnMQ2020}).
\item[(13)]
The determination of the \emph{smallest circle that encloses a set of $n$ static points and a mobile point}\/ $p$ in $\mathbb{R}^2$
(see, \emph{e.g.} Banik \emph{et al.} 2014 \cite{BanikBD2014}).
\item[(14)]
The localization of the \emph{center of the smallest enclosing circle of a set}\/ of $n$ points on a plane,
where the center is constrained to lie on a query line segment
(see, \emph{e.g.} Roy \emph{et al.} 2009 \cite{RoyKDN2009}).
\item[(15)]
The computation of the \emph{diameter}, also known as \emph{furthest pair}, of a set of $n$ points in $\mathbb{R}^d$
that is the maximum pair-wise Euclidean distance between two points in the set
(see, \emph{e.g.} Imanparast \emph{et al.} 2019 \cite{ImanparastHM2019}).
\item[(16)]
The approximation of the \emph{width} of a set of $n$ points in $\mathbb{R}^d$
that is the smallest distance of over all the regions between two parallel hyperplanes
that enclose all points of the set
(see, \emph{e.g.} Chan 2002 \cite{Chan2002}).
\item[(17)]
The determination of the \emph{flatness}\/ of a set of $n$ points in $\mathbb{R}^d$.
The mathematical notion of flatness is used in the
computational metrology scientific community and it corresponds exactly to the
width problem
(see, \emph{e.g.} Duncan \emph{et al.} 1997 \cite{DuncanGR1997}).
\item[(18)]
The measuring of the \emph{roundness}\/ of a set $P$ of $n$ points in $\mathbb{R}^d$
that can be determined by approximating $P$
with a sphere $S$ so that the maximum distance between a point of $P$ and $S$ is minimized.
This problem is equivalent to computing
the \emph{smallest closed region lying between two concentric spheres}\/ of radii $\varrho_1$ and $\varrho_2$ for $\varrho_1, \varrho_2 \in \mathbb{R}$ with
$ 0 \leqslant \varrho_1 \leqslant \varrho_2$ centered at $c \in \mathbb{R}^d$
that contain $P$
(see, \emph{e.g.} Agarwal \emph{et al.} 2000~\cite{AgarwalAHS2000}).
\item[(19)]
The \emph{covering}\/ of a set $P$ of $n$ points in $\mathbb{R}^2$
with a balanced V-shape of minimum width,
where a balanced V-shape is a polygonal region in the plane contained in the union of two
crossing equal-width strips
(see, \emph{e.g.} Aronov and Dulieu 2013~\cite{AronovD2013}).
\item[(20)]
The computation of the \emph{minimum (perimeter or area) axis-aligned rectangle enclosing $k$ points}\/
of a given set of $n$ points in the plane
(see, \emph{e.g.} Chan and Har-Peled 2021~\cite{ChanH2021}).
\item[(21)]
The treatment of the \emph{unsupervised $k$-windows clustering}\/
for improving the well-known and widely used
\emph{$k$-means clustering}\/
aiming at a better time complexity and partitioning accuracy,
and the determination of the number of clusters
(see, \emph{e.g.} Selim and Ismail 1984~\cite{SelimI1984},
Vattani 2011~\cite{Vattani2011},
Vrahatis \emph{et al.} 2002~\cite{VrahatisBAP2002}).
\item[(22)]
The computation of a \emph{minimum-width square or rectangular annulus}\/ that contains at least $n-k$ points out of $n$ given points in the plane,
where a square or rectangular annulus is the closed region between a square or rectangle and its offset.
The $k$ excluded points are considered as outliers of the $n$ input points
(see, \emph{e.g.} Bae 2019~\cite{Bae2019}).
\item[(23)]
The determination of the \emph{smallest enclosing cylinder}\/
that is the computation of the smallest radius over all cylinders that enclose
a set of $n$ points in $\mathbb{R}^d$, where a cylinder of radius $\varrho$ refers to the region of all points of
distance $\varrho$ from a line
(see, \emph{e.g.} Chan 2002 \cite{Chan2002}).
\item[(24)]
The computation of the \emph{width of simplices generated by the convex hull}\/ of their
integer vertices
(see, \emph{e.g.} Veselov \emph{et al.} 2019 \cite{VeselovGN2019}).
\item[(25)]
The determination whether a given \emph{simplex in $\mathbb{R}^d$ is covered by $d$-dimensional spheres}\/ centered at its vertices
(see, \emph{e.g.} Casado \emph{et al.} 2011 \cite{CasadoGTH2011}).
\item[(26)]
The determination of the \emph{largest {$k$}-dimensional ball in a {$d$}-dimensional box}\/
(see, \emph{e.g.} Everett \emph{et al.} 1998 \cite{EverettSVW1998}).
\hyphenation{com-pact}
\item[(27)]
The determination of the \emph{minimal volume of simplices containing a convex body}\/ $M$ in $\mathbb{R}^d$, where $M$ is a
compact convex set in $\mathbb{R}^d$ with non-empty interior
(see, \emph{e.g.}\ Kanazawa 2014~\cite{Kanazawa2014}, Galicer \emph{et al.} 2019 \cite{GalicerMF2019}).
\item[(28)]
The determination of the \emph{inscribing spheres in a body}\/ $M$ in $\mathbb{R}^d$ and \emph{inscribing $M$ in cylinders}, where $M$ is a
compact convex set in $\mathbb{R}^d$ with non-empty interior having $C^2$ smooth boundary (see, \emph{e.g.}\ Perel${}^\prime\!$man 1987~\cite{Perelman1987}).
\item[(29)]
The determination and
improvement
of \emph{bounds for outer (external) and inner (internal) radii quotient of a compact convex body in} $\mathbb{R}^d$
(see, \emph{e.g.}\ Pukhov 1979~\cite{Pukhov1979}, Perel${}^\prime\!$man 1987~\cite{Perelman1987},
Gonz\'{a}lez Merino 2017~\cite{Gonzalez-Merino2017}).
\end{itemize}

Theoretical issues for the
analysis and study of the MEB problem and the similar and related to it issues
can be found, among others, in the following fields
(appearing in alphabetical order):
(1)~Algorithmic Information Theory,
(2)~Applied Mathematics,
(3)~Approximation Algorithms Analysis,
(4)~Combinatorial Convexity,
(5)~Combinatorial Geometry,
(6)~Complex Projective Geometry,
(7)~Computational Biology,
(8)~Computational Complexity,
(9)~Computational Geometry,
(10)~Computational Information Geometry,
(11)~Computational Intelligence,
(12)~Computational Mathematics,
(13)~Convex Analysis,
(14)~Convex Geometry,
(15)~Convexity Theory,
(16)~Cryptography and Cryptanalysis,
(17)~Data Science,
(18)~Discrete Geometry,
(19)~Discrete Mathematics,
(20)~Geometric Computing,
(21)~Geometric Optimization,
(22)~Geometry and Topology,
(23)~Global Optimization,
(24)~Machine Intelligence and Learning,
(25)~Mathematical Programming,
(26)~Metric Geometry,
(27)~Non-Euclidean Geometry,
(28)~Operations Research,
(29)~Pattern Analysis,
(30)~Pure Mathematics,
(31)~Statistics,
(32)~Theoretical Computer Science,
(33)~Theory of Complexity,
(34)~Theory of Computation.

It is worth mentioning that, several theoretical approaches for the MEB and the related to it problems
have received a considerable impetus from their applications, that include, among others, the following ones:
\begin{itemize}
\leftskip0.05cm
\item[(1)]
Base station locations in facility locations (see, \emph{e.g.} Plastria 2002~\cite{Plastria2002}).
\item[(2)]
Detection of differential expression genes for RNA-seq data (see, \emph{e.g.} Zhou and Zhu~2022~\cite{ZhouZ2022}).
\item[(3)]
Roundness measurements in metrology (see, \emph{e.g.} Elerian \emph{et al.} 2021~\cite{ElerianHA2021}).
\item[(4)]
Gap tolerant classifiers in machine learning (see, \emph{e.g.} Burges 1998~\cite{Burges1998}).
\item[(5)]
$k$-center clustering (see, \emph{e.g.} B\u{a}doiu \emph{et al.} 2002~\cite{BadoiuHI2002}).
\item[(6)]
Solving the approximate 1-cylinder problem (see, \emph{e.g.} B\u{a}doiu \emph{et al.} 2002~\cite{BadoiuHI2002}).
\item[(7)]
Support vector clustering (see, \emph{e.g.} Ben-Hur \emph{et al.} 2001 \cite{Ben-HurHSV2001},
Bulatov \emph{et al.} 2004~\cite{BulatovJKS2004}).
\item[(8)]
Testing of clustering (see, \emph{e.g.} Alon \emph{et al.} 2000 \cite{AlonDPR2000},
Alon \emph{et al.} 2003 \cite{AlonDPR2003}, Alon \emph{et al.} 2004 \cite{AlonDPR2004}).
\item[(9)]
Support vector machine classification (see, \emph{e.g.} Rizwan \emph{et al.} 2021 \cite{RizwanIAK2021}).
\item[(10)]
Tuning support vector machine parameters (see, \emph{e.g.} Chapelle \emph{et al.} 2002 \cite{ChapelleVBM2002}).
\item[(11)]
Hand gesture recognition
(see, \emph{e.g}. Ren and Zhang 2009 \cite{RenZ2009}).
\item[(12)]
Preprocessing for fast farthest neighbor query approximation (see, \emph{e.g.} Goel \emph{et al.} 2001 \cite{GoelIV2001}).
\item[(13)]
Computation of spatial hierarchies, (\emph{e.g.} sphere trees) (see, \emph{e.g.} Hubbard 1996 \cite{Hubbard1996}).
\item[(14)]
Collision detection (see, \emph{e.g.} Hubbard 1996 \cite{Hubbard1996}).
\item[(15)]
Computer graphics
(see, \emph{e.g.} Larsson \emph{et al.} 2016 \cite{LarssonCK2016}).
\item[(16)]
The smallest covering cone (see, \emph{e.g.} Lawson 1965 \cite{Lawson1965}).
\item[(17)]
Performance of lightweight convolutional neural network models (see, \emph{e.g.} Tzelepi and Tefas 2019 \cite{TzelepiT2019}).
\item[(18)]
Similarity search in feature space (see, \emph{e.g.} Kurniawati \emph{et al.} 1997 \cite{KurniawatiJS1997}).
\item[(19)]
Fast fuzzy inference system training (see, \emph{e.g.} Chung \emph{et al.} 2009 \cite{ChungDW2009}).
\item[(20)]
Classification in non-stationary environments (see, \emph{e.g.} Heusinger and Schleif 2021 \cite{HeusingerS2021}).
\item[(21)]
Curve fitting for the case where a curve takes a sharp turn
(see, \emph{e.g.} Aronov and Dulieu 2013~\cite{AronovD2013}).
\item[(22)]
Minimum volume ball estimator in robust regression in statistics
(see, \emph{e.g.} Rousseeuw and Leroy 1987~\cite{RousseeuwL1987}).
\item[(23)]
Differential privacy (see, \emph{e.g.} Abowd 2018 \cite{Abowd2018}, Ghazi \emph{et al.} 2020 \cite{GhaziKM2020}, Mahpud and Sheffet 2022 \cite{MahpudS2022}).
\item[(24)]
Machine learning
(see, \emph{e.g.}
Tsang \emph{et al.} 2005 \cite{TsangKC2005},
Tsang \emph{et al.} 2006 \cite{TsangKZ2006},
Nielsen and Nock 2009 \cite{NielsenN2009},
Bauckhage \emph{et al.} 2019 \cite{BauckhageSD2019}).
\end{itemize}

\section{Upper Bounds of Radii of Convex Sets and Compact Convex Bodies}\label{sec:Relat}

\subsection{Relations and Upper Bounds of Circumradius, Inradius, Diameter and Width of Convex Sets}\label{subsec:Relat}

\hyphenation{in-equal-i-ties}
In 1958 Eggleston \cite{Eggleston1958}
considered all the twelve possible inequalities between any two of the four characteristics of a convex set $P$ in $\mathbb{R}^d$:
(a)
\emph{circumradius} denoted by $\rho_{\rm cir}^d (P)$,
(b)
\emph{inradius} denoted by $\rho_{\rm inr}^d (P)$,
(c)
\emph{diameter} denoted by ${\rm diam}(P)$ and
(d)
\emph{width} denoted by ${\rm wid}(P)$,
and categorized these inequalities as follows:
The first four obvious inequalities are the following:
\begin{subequations}\label{eggl-ineq:c1}
\begin{align}
\rho_{\rm inr}^d(P) &\leqslant \rho_{\rm cir}^d (P), & \rho_{\rm inr}^d(P)  & \leqslant \frac{1}{2} {\rm wid}(P), \label{eggl-ineq:c1a}\\[0.2cm]
{\rm diam}(P) &\leqslant 2 \rho_{\rm cir}^d (P), & {\rm wid}(P)  & \leqslant {\rm diam}(P). \label{eggl-ineq:c1b}
\end{align}
\end{subequations}
The above inequalities imply the following two ones:
\begin{equation}\label{eggl-ineq:c2}
\rho_{\rm inr}^d (P) \leqslant \frac{1}{2} {\rm diam}(P), \kern0.8cm
 {\rm wid}(P) \leqslant 2 \rho_{\rm cir}^d(P).
\end{equation}
In addition, Eggleston pointed out that since in the case where
the positive real number $\lambda \in \mathbb{R}^+$ is not infinite
there are no inequalities of the following four expressions:
\begin{subequations}\label{eggl-ineq:c3}
\begin{align}
\rho_{\rm cir}^d(P) &\leqslant \lambda\,\rho_{\rm inr}^d(P), & \rho_{\rm cir}^d(P)  & \leqslant \lambda\,{\rm wid}(P), \label{eggl-ineq:c3a}\\[0.2cm]
{\rm diam}(P) &\leqslant \lambda\,\rho_{\rm inr}^d(P), & {\rm diam}(P)  & \leqslant \lambda\,{\rm wid}(P), \label{eggl-ineq:c3b}
\end{align}
\end{subequations}
and the remaining two inequalities are very interesting and have the expressions:
\begin{equation}\label{eggl-ineq:c4}
\rho_{\rm cir}^d (P) \leqslant \lambda\,{\rm diam}(P), \kern0.8cm
{\rm wid}(P) \leqslant \lambda\,\rho_{\rm inr}^d (P).
\end{equation}
Particularly,
the first of the above inequalities~(\ref{eggl-ineq:c4})
is related to \emph{Jung's theorem}~\cite{Jung1901} (\emph{cf}.\ Theorem \ref{JungTh})
and it is given by
\begin{equation}\label{eggl-ineq:Jung}
\rho_{\rm cir}^d (P) \leqslant \sqrt{\frac{d}{2(d+1)}}\,\,{\rm diam}(P),
\end{equation}
while the second inequality
is related to
\emph{Steinhagen's theorem}~\cite{Steinhagen1921} (\emph{cf}.\ Theorem~\ref{SteinhagenTh})
and it is given by
\begin{equation}\label{eggl-ineq:Jungste}
{{\rm wid}(P)} \leqslant
\begin{cases}
{2 \sqrt{d}\, \rho_{\rm inr}^d (P)},&{\text{if}}\ {d}\,\ {\text{is odd}},\\[0.3cm]
\displaystyle{\frac{2(d+1)}{\sqrt{d+2}}\, \rho_{\rm inr}^d (P)},&{\text{if}}\ {d}\,\ {\text{is even}}.
\end{cases}
\end{equation}
\hyphenation{Reg-u-lar}
\subsection{Explicit Upper Bounds of Radii, Width and Median of Regular Simplices}

For a regular $d$-simplex $\tau^d$\/ in ${\mathbb{R}}^d$, if its diameter ${\rm diam}(\tau^d)$
(\emph{i.e}.\ the length of one of its equal edges) is given
then the circumradius $\rho_{\rm cir}^d (\tau^d)$, the inradius  $\rho_{\rm inr}^d (\tau^d)$ and the width ${\rm wid}(\tau^d)$ of $\tau^d$
can be explicitly determined in terms of ${\rm diam}(\tau^d)$ (\emph{cf}.\ Theorems~\ref{thm:circ-reg}, \ref{thm:inr-reg} and \ref{thm:reg-width}).
Also, a regular simplex is orthocentric,
its medians coincide with its corresponding altitudes and
the circumcenter $c_{\rm cir}^d$, the incenter $c_{\rm inr}^d$ and orthocenter $o^m$ coincide with the barycenter $\kappa^d$.
In addition,
the circumradius coincides with the barycentric circumradius $\beta_{\rm cir}^d (\tau^d)$, while the inradius
coincides with the barycentric inradius $\beta_{\rm inr}^m (\tau^d)$.

\begin{theorem}[Circumradius of regular simplices~\cite{Vrahatis2023}]\label{thm:circ-reg}
Let $\tau^d$ be a regular $d$-simplex in ${\mathbb{R}}^d$ with diameter
${\rm diam}(\tau^d)$,
then the circumradius $\rho_{\rm cir}^d (\tau^d)$ of $\sigma^d$\/ coincides with the barycentric circumradius $\beta_{\rm cir}^d (\tau^d)$ and
\begin{equation}\label{eq:equi-cir}
\rho_{\rm cir}^d (\tau^d) \equiv \beta_{\rm cir}^d (\tau^d) =
\sqrt{\frac{d}{2(d+1)}}\,\, {\rm diam}(\tau^d).
\end{equation}
\end{theorem}

\hyphenation{spher-i-cal}
Relation~(\ref{eq:equi-cir}) provides the radius of the
smallest spherical surface centered at the barycenter $\kappa^d$ enclosing the regular simplex $\tau^d$.
Note that the quantity in Relation~(\ref{eq:equi-cir}) constitutes the \emph{sharp upper bound}\/ of the inequality of the
significant \emph{Jung's enclosing theorem}\/ (\emph{cf}.\ Theorem~\ref{JungTh}).
\begin{theorem}[Inradius of regular simplices~\cite{Vrahatis2023}]\label{thm:inr-reg}
Let $\tau^d$ be a regular $d$-simplex in ${\mathbb{R}}^d$ with diameter
${\rm diam}(\tau^d)$,
then the inradius $\rho_{\rm inr}^d (\tau^d)$ of $\tau^d$\/ coincides with the barycentric inradius $\beta_{\rm inr}^d(\tau^d)$ and
\begin{equation}\label{eq:equi-inr}
\rho_{\rm inr}^d (\tau^d) \equiv \beta_{\rm inr}^d (\tau^d) =
\sqrt{\frac{1}{2 d (d+1)}}\,\, {\rm diam}(\tau^d).
\end{equation}
\end{theorem}
\begin{theorem}[Width of regular simplices~\cite{Vrahatis2023,Har-PeledR2023}]\label{thm:reg-width}
Let $\tau^d$ be a regular $d$-simplex in ${\mathbb{R}}^d$ with diameter
${\rm diam}(\tau^d)$,
then the width ${\rm wid}(\tau^d)$ of $\tau^d$ is given by:
\begin{equation}\label{eggl-ineq:Jungste-reg}
{{\rm wid}(\tau^d)} =
\begin{cases}
\displaystyle
{\sqrt{\frac{2}{d +1}}}\,\,{\rm diam}(\tau^d), &{\text{if}}\ {d}\,\ {\text{is odd}},\\[0.5cm]
\displaystyle
{\sqrt\frac{2 (d + 1)}{d (d + 2)}}\,\,{\rm diam}(\tau^d), &{\text{if}}\ {d}\,\ {\text{is even}}.
\end{cases}
\end{equation}
\end{theorem}

Next, we provide results related to the medians of regular simplices.

\begin{theorem}[Medians of regular simplices~\cite{Vrahatis2023}]\label{prop:med-reg}
Assume that $\tau^m = [{\upsilon}^0, {\upsilon}^1, \ldots, {\upsilon}^m]$
is a regular $m$-simplex in\/ ${\mathbb{R}}^d$ $(d \geqslant m)$
with edge length ${\rm diam}(\tau^m)$ and barycenter $\kappa^m$.
Let $\kappa_i^m$ be the barycenter of the
$i$-th $(m-1)$-face $\tau^{m}_{\neg i}$ of $\tau^m$
and let $\mu_i^m = [{\upsilon}^i,\, \kappa_i^m]$ be the $i$-th median that corresponds to the vertex ${\upsilon}^i$ for $i=0, 1, \ldots, m$.
Then
a) $\mu_i^m$ coincides with its corresponding $i$-th altitude,
b) $\tau^m$ is orthocentric, c) the orthocenter $o^m$ of $\tau^m$ coincides with its barycenter $\kappa^m$, and
d) it holds that:
\begin{equation}\label{eq:med-len-reg}
\|\mu_i^m\| = \sqrt{\frac{m+1}{2m}}\,\, {\rm diam}(\tau^m),\quad i=0, 1, \ldots, m.
\end{equation}
\end{theorem}
\begin{theorem}[Generalization of Pythagoras' theorem \cite{Vrahatis2023}]\label{prop:med-prop2}
Assume that $\tau^m = [{\upsilon}^0, {\upsilon}^1, \ldots, {\upsilon}^m]$
is a regular $m$-simplex in\/ ${\mathbb{R}}^d$ $(d \geqslant m)$
and
let $\kappa_i^m$ be the barycenter of the
$i$-th $(m-1)$-face $\tau^{m}_{\neg i}$ of $\tau^m$
for $i=0, 1, \ldots, m$.
Then, for $j\neq i$ and $i, j=0, 1, \ldots, m$, it holds that:
\begin{equation}\label{eq:med-prop2}
\|{\upsilon}^i -\kappa_i^m\|^2 + \|{\upsilon}^j -\kappa_{i}^m\|^2 = \|{\upsilon}^i -{\upsilon}^j\|^2.
\end{equation}
\end{theorem}

In general, the regular simplex exhibits interesting properties
including, among others, the \emph{maximizing properties} of the regular simplex
that have been given by Tanner in 1974~\cite{Tanner1974}.
Particularly, Tanner has proved
that the regular simplex maximizes the sum of the squared contents of all
$m$-faces, for all $m$, when the sum of squared edge lengths is fixed.
Thus, the regular simplex has the largest total length of all
edges, total area of all 2-faces,
total volume of all 3-faces,
\emph{etc.}, for a fixed sum of squared edge lengths~\cite{Tanner1974}.
Furthermore, in 1977 Alexander~\cite{Alexander1977} proved the conjecture of
Sallee\footnote{George Thomas Sallee (1940 -- 2019), American mathematician.}, which states that
\emph{``Only the regular simplex has maximal width among all simplices inscribed in a sphere of ${\mathbb{R}}^d$''},
(\emph{cf}.\ Theorem~\ref{thm:reg-width},
also see, \emph{e.g}.~\cite{Vrahatis2023},\cite{Alexander1977}, \cite{GritzmannK1992} and \cite{Har-PeledR2023}).

\subsection{Inner and Outer Radii of Compact Convex Bodies and Regular Polytopes}

Perel${}^\prime\!$man\footnote{Grigorii Yakovlevich Perel${}^\prime\!$man (b.\ 1966), Russian mathematician.}
in 1987 in his publication \emph{``$k$ radii of a compact convex body''}~\cite{Perelman1987} considered the internal and external $k$-radii
of a compact  (\emph{i.e}.\ closed and bounded) convex body $M$ in $\mathbb{R}^d$ that are particular cases of the
\emph{diameters}\/ used in \emph{{B}ern\v{s}te\u{\i}n and Kolmogorov approximation theory}, respectively.
Specifically, the \emph{internal (inner) $k$-radius}\/ $r_k(M)$ for $k = 1, 2, \ldots, d$ of $M$
is defined as the radius of the greatest $k$-dimensional sphere contained in $M$, while
the \emph{external (outer) $k$-radius}\/ $R_k(M)$ for $k = 1, 2, \ldots, d$ of $M$
is defined as the smallest radius of the solid cylinder containing $M$
with a $(d + 1 - k)$-dimensional spherical cross section and
a $(k - 1)$-dimensional generator.
Perel${}^\prime\!$man in~\cite{Perelman1987} and independently Pukhov in \cite{Pukhov1979} (\emph{cf.}~\cite{Gonzalez-Merino2017})
studied the
relation between these internal and external $k$-radii measures, and proved the following theorem:
\begin{theorem}[Perel${}^\prime\!$man-Pukhov theorem~\cite{Perelman1987}]\label{PerelPukhTh}
Let $M$ be a compact convex body in $\mathbb{R}^d$, then it holds that:
\begin{equation}\label{PerelPukhEq}
\frac{R_k(M)}{r_k(M)} \leqslant k + 1, \quad 1 \leqslant k \leqslant d ,
\end{equation}
where $R_k(M)$ and  $r_k(M)$ are the external and internal $k$-radii of $M$, respectively.
\end{theorem}

Also, Perel${}^\prime\!$man in~\cite{Perelman1987} pointed out that:
\begin{itemize}
\leftskip0.05cm
\item[(a)]
The problem of an upper bound for the radii quotient $R_{\ell}(M)/r_k(M)$ for $k \leqslant \ell$ which does not depend on
$M$ is interesting.
This quotient can be arbitrarily large for the case $k > \ell$.
\item[(b)]
Bounds which are, in fact, least upper bounds for $R_k(M)/r_k(M)$ is the most interesting, since the bound for $R_{\ell}(M)/r_k(M)$
for $k \leqslant \ell$  follows from it.
\item[(c)]
Least upper bounds for the cases $R_1(M)/r_1(M)$ and $R_d(M)/r_d(M)$ can be obtained
by Jung's theorem~\cite{Jung1901} and the generalization of Blaschke's theorem~\cite{Blaschke1914}
(due to Steinhagen \cite{Steinhagen1921}), respectively.
\item[(d)]
It is obvious that
$r_1(M) \geqslant r_2(M) \geqslant \cdots \geqslant r_d(M)$
and $R_1(M) \geqslant R_2(M) \geqslant \cdots \geqslant R_d(M)$.
Also, it holds that  $r_k(M) \leqslant R_k(M)$ for any $1 \leqslant k \leqslant d$.
\item[(e)]
The equality $r_k(M) = R_k(M)$ holds for certain bodies, \emph{e.g}.\ spheres.
\item[(f)]
Inequality~(\ref{PerelPukhEq}) is also valid for finite $r_k(M)$ without the assumption that the convex body $M$ is bounded.
\item[(g)]
The upper bound of Theorem~\ref{PerelPukhTh} is not a least upper bound
for any $1 \leqslant k \leqslant d$
and the estimate given in Inequality~(\ref{PerelPukhEq}) is extremely ``rough'' (non-sharp).
Also, precise estimates of $R_k(M)/r_k(M)$ for the cases $k = 2, 3, \ldots, d - 1$
are, evidently, unknown.
On the other hand, for the cases $k = 1$ and $k = d$,
the maximum of $R_k(M)/r_k(M)$ can be reached on a regular $d$-simplex.
\end{itemize}

Similarly, in 2017 Gonz\'{a}lez Merino~\cite{Gonzalez-Merino2017}
studied upper bounds for the quotient of successive
inner and outer radii of a convex body $M$ in $\mathbb{R}^d$.
In his approach the largest radius of an $i$-dimensional Euclidean disc contained in $M$
is denoted by $r_i(M)$, while the smallest radius of a solid cylinder
with $i$-dimensional spherical cross-section containing $M$ is denoted by $R_i(M)$, for
any $1 \leqslant i \leqslant d$. Using these notations the
\emph{Perel${}^\prime\!$man-Pukhov inequality}~(\ref{PerelPukhEq})
can be written as follows:
\begin{equation}\label{GonzEq1}
\frac{R_{d+1-i}(M)}{r_i(M)} \leqslant i + 1, \quad 1 \leqslant i \leqslant d ,
\end{equation}
Inequality~(\ref{GonzEq1}) extends two important results in convex geometry
that are particular cases of it.
Specifically, for $i = 1$ \emph{Jung's inequality}~\cite{Jung1901} is obtained, \emph{i.e}.
\begin{equation}\label{GonzEq2}
\frac{R_{d}(M)}{r_1(M)} \leqslant \sqrt{\frac{2 d}{d+1}} ,
\end{equation}
while, for $i = d$\/ \emph{Steinhagen's inequality}\/ \cite{Steinhagen1921} is attained, \emph{i.e}.
\begin{equation}\label{GonzEq3}
\frac{R_{1}(M)}{r_d(M)} \leqslant
\begin{cases}
\displaystyle
\sqrt{d}, &{\text{if}}\ {d}\,\ {\text{is odd}},\\[0.2cm]
\displaystyle
\frac{d+1}{\sqrt{d + 2}}, &{\text{if}}\ {d}\,\ {\text{is even}}.
\end{cases}
\end{equation}
Both Inequalities (\ref{GonzEq2}) and (\ref{GonzEq3}) are best possible,
since the regular $d$-simplex attains equality~\cite{Gonzalez-Merino2017}.

For the first time, in 2005 Brandenber~\cite{Brandenberg2005}
completely determined the inner and outer $j$-radii for three regular polytopes which exist in every dimension $d$,
namely, the \emph{regular simplex}, \emph{(hyper-) cube}, and \emph{regular cross-polytope}.
Brandenber defined the \emph{inner $j$-radii}\/ $r_j$\/ as the radii of the largest $j$-balls contained in $j$-dimensional slices of the polytope and
the \emph{outer $j$-radii}\/ $R_j$\/ as the radii of the smallest $j$-balls containing the projections of the polytope onto $j$-dimensional subspaces.
In general, the inner radii $r_j$\/ are also known as \emph{{B}ern\v{s}te\u{\i}n diameters}\/
and the outer radii $R_j$\/ as \emph{Kolmogorov diameters}\/ or \emph{Kolmogorov width}.
Among the results that have been presented in~\cite{Brandenberg2005}, the results of the radii of the three types of regular polytopes are exhibited below,
where the polytopes are scaled such that their circumradius to be 1.
Thus, (a) The inner radii $r_j$ of regular simplex are given as follows:
\begin{equation}\label{smRjSimEq}
r_j = \sqrt{\frac{d + 1}{j (j + 1) d}}.
\end{equation}
(b) The inner radii $r_j$ of both cube and regular cross-polytope are given by:
\begin{equation}\label{smRjCubeEq}
r_j = \sqrt{\frac{1}{j (d + 1)}}.
\end{equation}
(c) The outer radii $R_j$ of regular simplex are given as follows:
\begin{equation}\label{RjSimpEq}
R_j =
\begin{cases}
\displaystyle
\sqrt{\frac{j}{d}}, &{\text{if}}\ {j \notin \{1,\,d-1\}}\, \text{ or }\, d\, {\text{ odd}},\\[0.4cm]
\displaystyle
\frac{d+1}{d}\, \sqrt{\frac{1}{d + 2}}, &{\text{if}}\ {j =1}\, \text{and }\, d\, {\text{ even}},\\[0.4cm]
\displaystyle
\frac{2 d - 1}{2 d}, &{\text{if}}\ {j = d -1}\, \text{and }\, d\, {\text{ even}}.
\end{cases}
\end{equation}
(d) The outer radii $R_j$ of both cube and regular cross-polytope are given by:
\begin{equation}\label{RjCubeEq}
R_j = \sqrt{\frac{j}{d}} .
\end{equation}
Gritzmann and Klee in \cite{GritzmannK1992} showed that the computation of $R_j$ is NP-hard (nondeterministic polynomial-time hard)
for general simplices and many values of $j$.

\hyphenation{par-ti-tion-ing}
\section{Mathematical Approaches for Enclosing and Partitioning a Set}\label{sec:EnclTh}

\subsection{Theorems of Jung, Steinhagen and Perel${}^\prime\!$man}
In 1901 Jung gave seminal results regarding the MEB problem~\cite{Jung1901}.
Specifically, he gave results for the case of
\emph{finite point sets} and indicated their extension to \emph{infinite sets},
while in 1910 Jung~\cite{Jung1910} proposed necessary conditions on
the smallest circle enclosing a finite set of~$n$~points in a plane.
It is believed that the earliest known interesting application of Jung's theorem
was first given in 1905 by
Landau\footnote{Edmund Georg Hermann Landau (1877 -- 1938), German mathematician.}
\cite{Landau1905}
who applied Jung's theorem in a plane in order to sharpen an inequality related to {theory of analytic functions}.

\begin{theorem}[Jung's enclosing theorem (1901)~\cite{Jung1901}]\label{JungTh}
Assume that ${\rm diam}(P)$ is the diameter of a bounded subset
$P$ of\, $\mathbb{R}^d$ (containing more than a single point).
Then,
there exists a unique spherical surface of
circumradius $\rho_{\rm cir}^d (P)$ enclosing $P$, and it holds that
\begin{equation}\label{JungThEq}
\rho_{\rm cir}^d (P) \leqslant \sqrt{\frac{d}{2(d+1)}}\,\, {\rm diam}(P) .
\end{equation}
\end{theorem}

This significant Jung's enclosing theorem received quite a few additional proofs due to
S\"{u}ss 1936~\cite{Suss1936}, Blumenthal and Wahlin 1941~\cite{BlumenthalW1941},
Eggleston 1958~\cite[Th. 49, p. 111]{Eggleston1958} and
Guggenheimer 1977~\cite[Pr. 13-6, p. 140]{Guggenheimer1977}, among others.
Note that Jung's Inequality (\ref{JungThEq}) is \emph{sharp}\/ (best possible) since
the regular $d$-simplex attains equality (\emph{cf}.\ Theorem~\ref{thm:circ-reg}).

Jung's enclosing theorem is considered as a cornerstone in the field of MEB and related problems that
strongly influencing later developments and generalizations to other spaces and non-Euclidean geometries.
For example, to mention a few (appearing in chronological order):
\begin{itemize}
\leftskip0.2cm
\item[1938]
A generalization of Jung's theorem to \emph{Minkowski spaces}\/
has been proved by Bohnenblust~\cite{Bohnenblust1938}.
\item[1955]
A simpler proof of Bohnenblust's result~\cite{Bohnenblust1938} has been proposed by Leichtweiss~\cite{Leichtweiss1955}.
\item[1959]
The \emph{expansion constant}\/ and \emph{Jung's constant}\/ determined by the geometric properties
of a Minkowski space have been given by Gr\"{u}nbaum~\cite{Gruenbaum1959}.
\item[1984]
An \emph{infinity dimensional Hilbert space}\/ extension of Jung's theorem and its application to \emph{measures of noncompactness}\/
have been proposed by Dane\v{s}~\cite{Danes1984}.
\item[1985]
The determination of the \emph{Jung diameter}\/ of the \emph{complex projective plane}\/ has be given and
lower bounds for all \emph{complex projective spaces}\/ have been obtained by Katz~\cite{Katz1985}.
\item[1985]
A generalization of Jung's theorem has been proved by Dekster \cite{Dekster1985}
for which estimates are given for the side-lengths of certain inscribed simplices.
\item[1995]
The obtained result by Dekster \cite{Dekster1985} has been extended to the \emph{spherical and hyperbolic spaces}\/
by the same author~\cite{Dekster1995}.
\item[1997]
The result obtained in \cite{Dekster1985} extended by Dekster
to a class of \emph{metric spaces of curvature bounded above}\/
that includes \emph{Riemannian spaces}~\cite{Dekster1997}.
\item[1997]
An analogue of Jung's theorem for \emph{Alexandrov spaces of curvature bounded above}\/
has been presented in 1997 by Lang and Schroeder~\cite{LangS1997}.
\item[2006]
A complete characterization regarding
the \emph{extremal subsets of Hilbert spaces}, which constitutes
an infinite-dimensional generalization of the Jung theorem, has been proposed by
Nguen-Khac and Nguen-Van~\cite{Nguen-KhacN-V2006}.
\item[2006]
Generalizations of the Jung theorem for a \emph{pair of Minkowski spaces}\/
have been given by Boltyanski and Martini.
Particularly, the authors gave generalizations of Jung’s theorem for the case where there are two Minkowski
metrics in $\mathbb{R}^d$, one for the diameter of a set $P\in \mathbb{R}^d$, and the other for
the radius of the Minkowski ball containing $P$~\cite{BoltyanskiM2006}.
\item[2013]
\emph{Combinatorial generalizations}\/ of Jung's theorem
have been presented and
the \emph{``fractional''} as well as the \emph{``colorful''} versions of this theorem
have been proved by Akopyan~\cite{Akopyan2013}.
\item[2017]
An one-to-one connection between the \emph{Jung constant}\/ determined by Jung's theorem
in an arbitrary Minkowski space and the maximal \emph{Minkowski asymmetry}\/ of the
complete bodies within that space has been stated
by Brandenberg and Gonz\'{a}lez Merino~\cite{BrandenbergGM2017}.
\end{itemize}

It is worth reminding that,
Jung's theorem provides an upper bound of the circumradius of a set in terms of its diameter.
A similar important theorem due to Steinhagen provides lower bounds of the inradius of a set in terms of its width.
\begin{theorem}[Steinhagen's theorem (1921)~\cite{Steinhagen1921}]\label{SteinhagenTh}
Assume that $P$ is a bounded convex subset of\, $\mathbb{R}^d$ of minimal width ${\rm wid}(P)$.
Then, the inradius $\rho_{\rm inr}^d (P)$ satisfies:
\begin{equation*}\label{eq:steinh}
{\rho_{\rm inr}^d (P)} \geqslant
\begin{cases}
\bigl(2 \sqrt{d}\bigr)^{-1}\, {{\rm wid}(P)}, & {\text{if}}\ {d}\,\ {\text{is odd}},\\[0.2cm]
\bigl[\sqrt{d+2}/(2d+2)\bigr]\, {{\rm wid}(P)}, & {\text{if}}\ {d}\,\ {\text{is even}}.
\end{cases}
\end{equation*}
\end{theorem}

The Steinhagen theorem has been proved in 1914 for $d = 2$ by Blaschke~\cite{Blaschke1914}.
An additional proof of Steinhagen's theorem has been given in 1936 by Gericke~\cite{Gericke1936}.
Also, Santal\'{o} in 1946~\cite{Santalo1946}
gave a \emph{generalization of Jung's and Steinhagen's theorems} to convex regions on the \emph{$d$-dimensional spherical surface}.
A geometrical proof of the above results for $d = 2$  has been given in 1944, also by Santal\'{o}~\cite{Santalo1944}.
Santal\'{o}'s result about the generalization of the Steinhagen's theorem, also generalizes
the well-known \emph{Robinson's theorem} that it has been proved by Robinson in 1938~\cite{Robinson1938}.
In 1992 Henk~\cite{Henk1992} by studying the relation between circumradius and diameter of a convex
body which has been proposed by Jung's theorem, gave
a natural \emph{generalization of Jung's theorem}.
Similarly, in 1993 Betke and Henk~\cite{BetkeH1993} by analyzing the relation between inradius and
width of a convex set that has been established by Steinhagen's theorem proposed respectively
a natural \emph{generalization of Steinhagen's theorem}.

Perel${}^\prime\!$man in 1987~\cite{Perelman1987}
applied the concept of $k$ radii (\emph{cf}.\ Sec.~\ref{sec:Relat}) for the solution of two problems for compact convex
bodies with $C^2$ smooth boundaries that
have been asked by Ionin
at the geometric seminar of the Novosibirsk Mathematics Institute in 1984
and are referred as Problems 8 and 9.
The problems are
related to inscribing spheres in a body $M$ in $\mathbb{R}^d$ and inscribing $M$ in cylinders, where $M$ is a
compact convex body having $C^2$ smooth boundary.
Specifically, Perel${}^\prime$man proved the following theorem:

\begin{theorem}[Perel${}^\prime\!$man's enclosing theorem (1987)~\cite{Perelman1987}]\label{PerelmanTh}
Let $M$ be a compact convex body in $\mathbb{R}^d$ with a $C^2$ smooth boundary. If there are $k$ principal curvatures not greater than~1 at each point of the boundary of $M$ (where the number $k$ is the same for all points of the boundary), then it is possible to place in $M$ a $(k + 1)$-dimensional ball of radius not less than $1/(k - 2)$. Correspondingly, if the $k$ principal
curvatures are not less than 1 at any point of the boundary, then $M$ can be enclosed in a solid cylinder with a $(k + 1)$-dimensional spherical cross section of radius not greater than
1 and a $(d - k - 1)$-dimensional generator for any $1 \leqslant k \leqslant d-1$.
\end{theorem}

The following result permits the reduction of the MEB problem to
a finite one concerning $d+1$ points.

\begin{theorem}[Blumenthal-Wahlin enclosing theorem (1941) \cite{BlumenthalW1941}]\label{lem:Car}
If each set of\/ $d+1$ points of a subset $P$ of\/ $\mathbb{R}^d$ is enclosable by a spherical surface of a given radius,
then $P$ is itself enclosable by this spherical surface.
\end{theorem}

In addition, in 1953 Gale~\cite{Gale1953} gave
the following  equivalent formulation of Jung's enclosing theorem.

\begin{theorem}[Gale's enclosing theorem (1953)~\cite{Gale1953}]\label{GaleTh-cir}
The circumscribed $d$-sphere of a regular $d$-simplex of diameter~1 will cover any $d$-dimensional set of diameter~1.
\end{theorem}

Next, we provide useful results for simplices.
To this end, the following notation and assumptions are used.

\begin{notation}\label{nota:edges}
The sum of the squares of the lengths of
the $m$ edges of an $m$-simplex in\/ ${\mathbb{R}}^d$ $(d \geqslant m)$
$\sigma^m = [{\upsilon}^0, {\upsilon}^1, \ldots, {\upsilon}^m]$
that concur in the vertex ${\upsilon}^i$ for $i=0,1, \ldots, m$\/
is denoted by:
\begin{equation}\label{eq:ei}
E({\upsilon}^i)=\sum_{\substack{j=0\\j\ne i}}^m\|{\upsilon}^i - {\upsilon}^j\|^2.
\end{equation}
Furthermore, the sum of the squares of the lengths of
the $\binom{m}{2}= (m-1)m/2$\, edges of the $i$-th $(m-1)$-face $\sigma^{m}_{\neg i}$ of $\sigma^m$
is denoted as follows:
\begin{equation}\label{eq:enegi}
E(\sigma_{\neg i}^m)=
\sum_{\substack{p=0\\p\ne i}}^{m-1}\,
\sum_{\substack{q=p+1\\q\ne i}}^m\|{\upsilon}^p - {\upsilon}^q\|^2.
\end{equation}
In addition, the sum of the squares of the lengths of all
the $\binom{m+1}{2}= m(m+1)/2$\, edges of $\sigma^m$
is denoted by:
\begin{equation}\label{eq:enall}
E(\sigma^m)=
\sum_{p=0}^{m-1} \sum_{q=p+1}^{m}\| {\upsilon}^p - {\upsilon}^q \|^2.
\end{equation}

\end{notation}

\begin{assumptions}\label{assum:1}
Suppose that $\sigma^m = [{\upsilon}^0, {\upsilon}^1, \ldots, {\upsilon}^m]$
is an $m$-simplex in\/ ${\mathbb{R}}^d$ $(d \geqslant m)$
with barycenter $\kappa^m$ and diameter ${\rm diam}(\sigma^m)$.
Let $\kappa_i^m$ be the barycenter of the
$i$-th $(m-1)$-face $\sigma^{m}_{\neg i}$ of $\sigma^m$ and
let $\mu_i^m = [{\upsilon}^i,\, \kappa_i^m]$ be the $i$-th median that corresponds to the vertex ${\upsilon}^i$ for $i=0, 1, \ldots, m$.
\end{assumptions}

\begin{theorem}[Generalization of Apollonius' theorem \cite{Vrahatis2023}]\label{th:apol}
Let Assumptions~\ref{assum:1} hold.
Then,
\begin{equation}\label{eq:med-len}
\|\mu_i^m\|^2 = \frac{1}{m^2} \bigl( m E({\upsilon}^i) - E(\sigma_{\neg i}^m) \bigr), \quad  i=0, 1, \ldots, m.
\end{equation}
\end{theorem}

\begin{theorem}[Generalization of Commandino's theorem \cite{Vrahatis2023}]\label{SimCommandinoTh}
Let Assumptions~\ref{assum:1} hold.
Then, the $m+1$ medians of
$\sigma^m$ concur in the barycenter $\kappa^m$ of $\sigma^m$
that divides each of them in the ratio $1:m$, where the longer segment being on
the side of the vertex of $\sigma^m$, that is to say,
\begin{equation}\label{eq:Km-rat}
\| \kappa^{m} - {\upsilon}^i \| = m \| \kappa^{m} - \kappa^{m}_{i} \|, \quad i=0, 1, \ldots, m.
\end{equation}
\end{theorem}

\begin{theorem}[Barycentric distances of the vertices~\cite{Vrahatis1986,Vrahatis2023}]\label{prop:bar-prop3}
Let Assumptions~\ref{assum:1} hold.
Then for each $i=0, 1, \ldots, m$\/ it holds that:
\begin{equation}\label{eq:bar-prop3}
\| \kappa^{m} - {\upsilon}^i \| = \frac{1}{m+1}\,\,\sqrt{ m E({\upsilon}^i) - E(\sigma_{\neg i}^m)}.
\end{equation}
\end{theorem}

\begin{theorem}[Barycentric sum property~\cite{Vrahatis2023}]\label{prop:bar-prop}
Let Assumptions~\ref{assum:1} hold.
Then,
\begin{equation}\label{eq:bar-prop}
\sum_{i=0}^{m} \| \kappa^{m} - {\upsilon}^i \|^2 = \frac{1}{m+1}\, E(\sigma^m).
\end{equation}
\end{theorem}

\begin{theorem}[Medians sum property~\cite{Vrahatis2023}]\label{prop:med-prop}
Let Assumptions~\ref{assum:1} hold.
Then,
\begin{equation}\label{eq:med-prop}
\sum_{i=0}^{m} \| \mu_i^m \|^2 = \frac{m+1}{m^2} E(\sigma^m).
\end{equation}
\end{theorem}

\begin{theorem}[Simplex barycentric enclosing~\cite{Vrahatis2023}]\label{thm:simcov}
Let Assumptions~\ref{assum:1} hold.
 Then the barycentric circumradius $\beta_{\rm cir}^m$ is unique, it is given as follows:
\begin{equation}\label{eq:3.3}
\beta_{\rm cir}^m = \frac{1}{m + 1} \max_{0 \leqslant i \leqslant m} \bigl\{ m E({\upsilon}^i) - E(\sigma_{\neg i}^m) \bigr\}^{1/2},
\end{equation}
and the spherical surface with center at $\kappa^m$ and radius $\beta_{\rm cir}^m$ encloses $\sigma^m$.
\end{theorem}

\begin{theorem}[Simplex enclosing~\cite{Vrahatis2023}]\label{thm:simcovII}
Let Assumptions~\ref{assum:1} hold.
Suppose further that
$\beta_{\rm cir}^m$ is the barycentric circumradius.
Then, there exists a unique spherical surface of
circumradius $\rho_{\rm cir}^m$ enclosing $\sigma^m$, and
\begin{equation}\label{eq:simcovrho}
\rho_{\rm cir}^m \leqslant \min \bigl\{\beta_{\rm cir}^m,\, [m/(2m+2)]^{1/2}\, {\rm diam}(\sigma^m) \bigr\}.
\end{equation}
\end{theorem}

The computation of $\rho_{\rm cir}^m$ does not require
any additional computational burden since it can be easily obtained
during the computation of the longest edge (diameter) of the simplex.

Note that, the barycentric circumradius $\beta_{\rm cir}^d=\beta_{\rm cir}^d(P)$ of a set $P$ in $\mathbb{R}^d$
is the supremum of the barycentric circumradii of simplices
with vertices in $P$.
Next, it has been shown in Theorem~\ref{VrahatisTheoremScov}, that a bounded set $P$ can be covered by a spherical surface
of circumradius~$\beta_{\rm cir}^d(P)$,
which in many cases gives a better result than Jung's theorem (\emph{cf.} \cite{Vrahatis1988}).
For instance, consider
a regular $2$-simplex $\tau^2 = [{\upsilon}^0, {\upsilon}^1, {\upsilon}^2]$ with ${\rm diam}(\tau^2) =1$
and barycenter $\kappa^2$.
By  Theorem~\ref{thm:circ-reg}
we obtain for the circumradius of $\tau^2$,
$\rho_{\rm cir}^2(\tau^2) = 3^{-1/2}$
that is the upper bound given by
Jung's Theorem~\ref{JungTh}.
Consider now the $2$-simplex $\sigma^2 = [\kappa^2, {\upsilon}^1, {\upsilon}^2]$.
The lengths of the edges of $\sigma^2$ are $3^{-1/2}$, $3^{-1/2}$ and 1.
Thus, ${\rm diam}(\sigma^2) =1$ and consequently $\rho_{\rm cir}^2(\sigma^2) = 3^{-1/2}$.
By Theorem~\ref{thm:simcov} we obtain for the
barycentric circumradius of $\sigma^2$, $\beta_{\rm cir}^d(\sigma^2) = [7/27]^{1/2}$
which is smaller than the upper bound $3^{-1/2}$ for
the circumradius $\rho_{\rm cir}^2$ that is given by
Jung's covering Theorem~\ref{JungTh}.

\begin{theorem}[Variant of Jung's theorem (1988)~\cite{Vrahatis1988}]\label{VrahatisTheoremScov}
Assume that ${\rm diam}(P)$ is the diameter of a bounded subset
${P}$ of\, $\mathbb{R}^d$ (containing more than a single point)
and let $\beta_{\rm cir}^d=\beta_{\rm cir}^d(P)$ be its barycentric circumradius.
Then, there exists a unique spherical surface of
circumradius $\rho_{\rm cir}^d$ enclosing $P$, and
\begin{equation}\label{eq:pcovrho}
\rho_{\rm cir}^d \leqslant \min \bigl\{\beta_{\rm cir}^d(P),\, [d/(2d+2)]^{1/2}\, {\rm diam}(P) \bigr\}.
\end{equation}
\end{theorem}

\subsection{Theorems of Carath\'{e}odory, Helly, Radon and Tverberg}

Carath\'{e}odory\footnote{Constantin Carath\'{e}odory (1873 -- 1950), Greek mathematician.}
in 1907~\cite{Caratheodory1907}
gave an important theorem in convex geometry.
In 1970
Rockafellar\footnote{Ralph Tyrrell Rockafellar (b.\ 1935), American mathematician.}
\cite{Rockafellar1970} has pointed out that
\emph{``Carath\'{e}odory's theorem is the fundamental dimensionality result in
convexity theory, and it is the source of many other results in which
dimensionality is prominent.''}
In general, the convex hull of a subset $P$ of $\mathbb{R}^d$ can be obtained
by carrying out all convex combinations of elements of $P$.
On the other hand, due to Carath\'{e}odory's theorem
it is not necessary to perform combinations involving
more than $d + 1$ elements at a time.
Thus, convex combinations of the form $\sum_{i=1}^{m} \lambda_i x_i$
can be performed where $m \leqslant d +1$
or $m=d + 1$
in the case of non-distinct vectors~$x_i$ (\emph{cf}.~\cite{Rockafellar1970}).
\emph{Carath\'{e}odory's theorem in convexity theory}\/ states that:

\begin{theorem}[Carath\'{e}odory's theorem (1907)~\cite{Caratheodory1907}]\label{CaratheodoryTh}
Any point in the convex hull of a finite point set in $\mathbb{R}^d$ is a convex combination
of some at most $d + 1$ of these points.
\end{theorem}

Carath\'{e}odory's theorem is related to the following partitioning theorems
due to
Helly\footnote{Eduard Helly (1884 -- 1943), Austrian mathematician.}
and
Tverberg\footnote{Helge Arnulf Tverberg (1935 -- 2020), Norwegian mathematician.}.

\begin{theorem}[Helly's partitioning theorem (1913)~\cite{Helly1923}]\label{HellyTh}
Let $C_1, C_2, \ldots, C_k$ be a finite family of convex subsets of\/ $\mathbb{R}^d$, with $k \geqslant d + 1$.
If the intersection of every $d + 1$ of these sets is nonempty, then the whole family has a nonempty intersection,
\emph{i.e.}~$\cap_{i=1}^{k} C_i \neq \emptyset$.
\end{theorem}

\begin{theorem}[Tverberg's partitioning theorem (1966)~\cite{Tverberg1966}]\label{TverbergTh}
Every set with at least $(p - 1)(d + 1) + 1$ points in\/ $\mathbb{R}^d$ can be
partitioned into $p$ subsets whose convex hulls all have at least one point in common.
\end{theorem}

Theorems~\ref{CaratheodoryTh}, \ref{HellyTh} and \ref{TverbergTh}
are equivalent in the sense that each one can be deduced from another~\cite{DeLoeraGMM2019}.
Theorem~\ref{HellyTh} proposed by Helly in 1913 \cite{DanzerGK1963} but not published by him until 1923.
Tverberg gave the first proof of Theorem~\ref{TverbergTh} in 1966~\cite{Tverberg1966},
while in 1981 he gave a simpler proof~\cite{Tverberg1981}.
Tverberg's partitioning theorem constitutes  a generalization of a theorem due to
Radon\footnote{Johann Karl August Radon (1887 -- 1956), Austrian mathematician.}
that it has been proposed in 1921:

\begin{theorem}[Radon's partitioning theorem (1921)~\cite{Radon1921}]\label{RadonTh}
Every set with $d + 2$ points in $\mathbb{R}^d$ can be partitioned into two sets whose convex hulls intersect.
\end{theorem}

Theorems~\ref{CaratheodoryTh}, \ref{HellyTh}, \ref{TverbergTh} and \ref{RadonTh} are known as
\emph{Helly-type theorems}~\cite{BaranyK2022}.
In addition, a classical partitioning theorem for points in the plane that is also based on Carath\'{e}odory's theorem
has been proved by
Birch\footnote{Bryan John Birch (b. 1931), English mathematician.}
in 1959~\cite{Birch1959}.
Also, Adiprasito \emph{et al}.\ in 2020 \cite{AdiprasitoBMT2020}
initiated the study of the dimensionless versions
of classical theorems in convexity theory.
Specifically, they considered the dimensionless versions of the
theorems of Carath\'{e}odory, Helly, and Tverberg.
The obtained results have several ``colorful'' and ``fractional'' consequences
and are particulary interesting and motivating, among others, for those who are interested in classical convexity
parameters. The authors named these theorems as \emph{``no-dimension theorems''}\/ and proved the following theorems, among others:

\begin{theorem}[No-dimension Carath\'{e}odory theorem~\cite{AdiprasitoBMT2020}]\label{NoColCarathTh}
Let $P$ be a set of $n$ points in $\mathbb{R}^d$, $r \in [n]$ (\emph{i.e}.\ $r \leqslant n$), and $a \in {\rm conv}\,\!P$.
Then there exists a subset $Q$ of $P$ with $r$ elements (\emph{i.e}.\ $|Q| = r$) such that for the distance between
$a$ and the convex hull of $Q$ holds that:
\begin{equation*}\label{NoColCarathEq}
{\rm dist} \bigl(a,\, {\rm conv}\,Q\bigr) \leqslant \frac{{\rm diam}(P)}{\sqrt{2 r}} .
\end{equation*}
\end{theorem}

\begin{theorem}[No-dimension Helly theorem~\cite{AdiprasitoBMT2020}]\label{NoColHellyTh}
Assume that $K_1, K_2, \ldots,K_n$ are convex sets in $\mathbb{R}^d$ and $k \in [n]$. For $J \subset [n]$
define $K(J ) = \bigcap_{j \in J} K_j$. If the Euclidean unit ball $B(b, 1)$ centered at $b \in \mathbb{R}^d$
intersects $K(J)$ for every $J \subset [n]$ with $|J| = k$, then there is a point $q \in \mathbb{R}^d$ such
that
\begin{equation*}\label{NoColHellyEq}
{\rm dist} \bigl(q,\, K_i\bigr) < \frac{1}{\sqrt{k}}, \quad \forall\, i\in [n].
\end{equation*}
\end{theorem}

\begin{theorem}[No-dimension Tverberg theorem~\cite{AdiprasitoBMT2020}]\label{NoColTverbTh}
Let $P$ be a set of $n$ points in $\mathbb{R}^d$,
then for a given integer $1 \leqslant k \leqslant n$,
there exists
a point $q \in \mathbb{R}^d$ and a partition of $P$ into $k$ sets $P_1, P_2, \ldots, P_k$ such that:
\begin{equation*}\label{NoColTverbEq}
{\rm dist} \bigl(q,\, {\rm conv}\,\!P_i\bigr) \leqslant \bigl(2 + \sqrt{2}\bigr)\, \sqrt{ \frac{k}{n}}\,\, {\rm diam}(P),  \quad \forall\, i\in [k].
\end{equation*}
\end{theorem}

Carath\'{e}odory's theorem has obtained numerous interesting reformulations, variations, generalizations and applications
(see, \emph{e.g}.\ Danzer \emph{et al}.\ 1963 \cite{DanzerGK1963},
De Loera \emph{et al}.\ 2019 \cite{DeLoeraGMM2019}).
Rockafellar in 1970~\cite{Rockafellar1970}
used Carath\'{e}odory's theorem to prove Helly's theorem
and results related to \emph{infinite systems of linear inequalities}.
In addition, various generalizations, refinements, applications and
conjectures of some of the classical combinatorial
theorems about convex sets (Carathéodory, Helly, Radon and Tverberg theorems)
have been published in 2022 by B\'{a}r\'{a}ny and Kalai~\cite{BaranyK2022}.
Also, in~\cite{BaranyK2022} the authors point out the connection between important results
  from combinatorial convexity and some theorems from topology. For instance, Helly's theorem
  is a manifestation of the \emph{nerve theorem}\/ from algebraic topology,
 and Radon’s theorem can be regarded as an early \emph{linear}\/ version
 of the \emph{Borsuk-Ulam theorem}.

In 1982 B\'{a}r\'{a}ny \cite{Barany1982} proposed the following
sharp generalization of Carath\'{e}odory's theorem
in the sense that the number of partition of the sets $C_i$'s cannot be decreased.

\begin{theorem}[B\'{a}r\'{a}ny's partitioning theorem (1982)~\cite{Barany1982}]\label{BaranyTh}
Let $C_0, C_1, \ldots, C_d$\, be subsets of\/ $\mathbb{R}^d$
and let $\alpha$ be a point in ${\rm conv}\,C_i$ (the convex hull of $C_i$)
for $i=0,1,\ldots, d$ (\emph{i.e}.\ $\alpha \in \bigcap_{i=0}^d {\rm conv}\,C_i$).
Then there exist vectors ${\upsilon}^i \in C_i$ for $i=0,1,\ldots, d$\,
such that $\alpha \in {\rm conv} \{{\upsilon}^0, {\upsilon}^1, \ldots, {\upsilon}^d \}$.
\end{theorem}

In addition, in a paper published in 1986 by Polikanova and Perel${}^\prime$man,
titled \emph{``A remark on {H}elly's theorem''},
the following partitioning theorem has been proved~\cite{PolikanovaP1986}:

\begin{theorem}[Polikanova-Perel${}^\prime\!$man partitioning theorem (1986)~\cite{PolikanovaP1986}]\label{PolikanovaPTh}
Let $\mathfrak{M}$ be a bounded family of compact convex sets in $\mathbb{R}^d$
such that $\mu_d(\bigcap \mathfrak{M})=0$, where $\mu_d$ denotes the $d$-dimensional volume.
Then for every $\varepsilon > 0$ there are $d+1$ sets
$M_1, M_2, \ldots, M_{d+1}$ in the family $\mathfrak{M}$ such that
$\mu_d \bigl(\bigcap_{i=1}^{d+1} M_i \bigr) < \varepsilon$.
Also, if $\dim \bigcap \mathfrak{M} = \ell$,\, $0\leqslant \ell < d$,
then there are $d-\ell+1$ sets in the family $\mathfrak{M}$
such that $\mu_d \bigl(\bigcap_{i=1}^{d -\ell +1} F_i \bigr) \leqslant \varepsilon$.
\end{theorem}

\section{Concluding Remarks}\label{sec:Synop}
A characteristic of the MEB and related to it problems is that the theory and applications
of this field of study and research are very broad and divers.
Also, a large variety of methods and algorithms
for approximating the MEB, diameter and width of a set
have been proposed in the literature.

Our contribution is focused mainly on essentials
enclosing and partitioning theorems
that are considered as cornerstones in the
field that strongly influencing developments and generalizations
to other spaces and non-Euclidean geometries.
Specifically, a contribution is provided towards the mathematical foundation
of a framework to study, analyze and apply rigorously
the MEB and relared to it problems.

We hopefully think that the paper at hand might contribute
to a deeper understanding and rigorous tackling of
the MEB problem and
the similar and related to it issues and aspects.

\vfill
\begin{IEEEbiography}[{\includegraphics[width=1in,height=1.25in,clip,keepaspectratio]{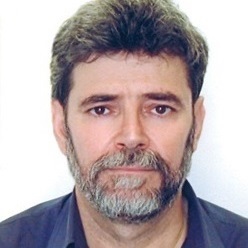}}]{Michael N. Vrahatis}
is a professor emeritus 
of the University of Patras, Greece.
His main scientific objectives are:
(a) the investigation and mathematical foundation of methods in artificial intelligence,
(b)~the development of innovative methods in this field,
(c) the application of the acquired expertise to address challenging real-life problems in different branches of science and
the search of directions of theoretical mathematical results towards applications, and
(d) the dissemination of knowledge
to young researchers and scholars. 
His current 
interests focus on the thematic areas:
(a) mathematics,
(b) natural computing and computational intelligence,
(c) global optimization,
(d) reliable computing and imprecise data, and
(e) neural networks and machine learning~\& intelligence. 
\end{IEEEbiography}
\end{document}